\documentclass[prb,twocolumn,showpacs,preprintnumbers,amsmath,amssymb,superscriptaddress,floatfix]{revtex4-1}
\usepackage{graphicx}
\usepackage{color}
\usepackage{dcolumn}
\usepackage{bm}
\usepackage[latin1]{inputenc}
\def\IR{\relax{\rm I\kern-.18 em R}}

\begin{document}
\title{Inversion mechanism for the transport current in type-II superconductors}
\author{H. S. Ruiz}
\email[Electronic address: ]{hsruizr@unizar.es}
\affiliation{Departamento de F\'{\i}sica de la Materia
Condensada--I.C.M.A., Universidad de Zaragoza--C.S.I.C., Mar\'{\i}a
de Luna 1, E-50018 Zaragoza, Spain}

\author{C. L\'opez}
\affiliation{Departamento de Matem\'aticas, Universidad de Alcal\'a
de Henares, E-28871 Alcal\'a de Henares, Spain}

\author{A. Bad\'{\i}a\,--\,Maj\'os}
\affiliation{Departamento de F\'{\i}sica de la Materia
Condensada--I.C.M.A., Universidad de Zaragoza--C.S.I.C., Mar\'{\i}a
de Luna 1, E-50018 Zaragoza, Spain}

\date{\today}

\begin{abstract}

The longitudinal transport problem (current is applied parallel to some bias magnetic field) in type-II superconductors is analyzed theoretically. Based on analytical results for simplified configurations and relying on numerical studies for general scenarios, it is shown that a remarkable inversion of the current flow in a surface layer may be predicted under a wide set of experimental conditions. Strongly inhomogeneous current density profiles, characterized by {\em enhanced} transport towards the center and reduced or even negative values at the periphery of the conductor are expected when 
the physical mechanisms of flux depinning and consumption (via line cutting) are recalled. A number of striking collateral effects such as local and global paramagnetic behavior are predicted. Our geometrical description of the macroscopic material laws allows a pictorial interpretation of the physical phenomena underlying the transport backflow. 

\end{abstract}
\maketitle

\section{Introduction}

Type-II superconductors under the action of a transport current and a longitudinal magnetic field may exhibit the counter intuitive phenomenon of negative resistance within a certain set of experimental conditions. This property, together with other intriguing phenomena, such as the observation of paramagnetic moments, and the {\em compression} of the transport current by the action of a parallel magnetic field, have been reported in the course of intense experimental and theoretical activities.\cite{parallelE,walmsley,cave,matsushita,parallelT,voloshin,fisher} Most of these works were primarily concerned with the arrangement of the macroscopic current density ${\bf J}$ along the so-called nearly {\em force free} trajectories. Recall that if ${\bf J}$ is {\em nearly parallel} to the magnetic induction ${\bf B}$, moderate or weak pinning forces are needed for avoiding the detrimental flux-flow losses related to the drift of flux tubes driven by the magnetostatic force (${\bf J}\times{\bf B}$ per unit volume). More specifically, negative voltages have been observed by different groups\cite{walmsley,cave,matsushita} when recording the current-voltage characteristics at specific locations on the surface of the sample (central region). In addition, the striking effect takes place within a definite interval of applied magnetic fields.

Within such a complex scenario, it was recognized early on that the observations could only be understood if new dissipation mechanisms, additional to the flux flow phenomena were considered. In particular, a prominent role happens to be played by the flux-line cutting (crossing and recombination) between adjacent tilted vortices.\cite{parallelT,leblanc} Nevertheless, certain facts still remain to be fully understood. Thus, the challenging problem of inhomogeneous electric fields, even changing sign along the specimen surface is hitherto open. On the other hand, important issues as the consideration of irreversible effects related to the thresholds for flux depinning and cutting phenomena have not been reported yet.
In this contribution, we investigate the influence of such mechanisms on the establishment of critical negative current structures within the superconducting state. This is to be considered as a step forward for gaining knowledge on the processes that operate just previous to the dissipation regime. Within such a physical scenario, the concentration of transport current towards the center of  the sample and the appearance of negative flow at the surface will be predicted for a certain range of experimental conditions. To be specific, the application of different components of magnetic field, their sequence and characteristic values will be identified as relevant issues for the observation of negative currents.

The article is organized as follows. In Sec.\ref{secCS} we put forward the basic ideas about the theoretical approach used, that is a general critical state theory for type-II superconductors. An idealized slab geometry arrangement is proposed, aimed to introduce the lowest level of complexity for our purposes. Then, in Sec.\ref{secAnalytic} we perform a simplified analytical evaluation that allows to capture the main underlying physical matters. In Sec.\ref{secNumeric} a quasi-3D statement of the problem is solved by numerical means. This is needed for the consideration of inhomogeneities over the sample. A discussion about the scope of our investigation within the problem of negative currents is given in Sec.\ref{secDisc} 

\section{Critical state approximation}
\label{secCS}

\subsection{Classical Maxwell equation approach}

The fundamental  concept on which the critical state theory relies is that, in many cases, the experimental conditions allow to analyze the evolution of the system in the quasistationary regime. Thus, Ampere's law becomes $\nabla \times  {\bf H} = {\bf J}$, and determines the distribution of supercurrents within the sample. When some external excitation (magnetic field and/or transport current) applies, the quasistationary evolution between successive equilibrium states is ruled by Faraday's law $\nabla \times {\bf E} = - \partial _t {\bf B}$. Here the induced transient electric field is determined through an appropriate material relation ${\bf J}({\bf E})$, and is used to update the profile of $\bf J$.

The material law encodes the mechanisms related to the breakdown of magnetostatic equilibrium, as well as the dissipation modes operating in the transient from one state to the other. In this sense, there have been a number of theoretical proposals, and among them (see Refs.\onlinecite{prl,gcs} and the citations therein), here we choose the so-called {\em double critical state model\,} (DCSM).\cite{dcsm}  This approach allows a straightforward connection between the mesoscopic flux depinning and cutting phenomena, and the field equations for the coarse grained quantity $\bf J$. On the one hand, the model establishes the critical conditions $|{\bf J}_{\perp}|\leq J_{c\perp}$ and $|{\bf J}_{\parallel}|\leq J_{c\parallel}$ that relate to (i) the maximum pinning force on the vortex lines ($|{\bf J}\times{\bf B}|=  J_{\perp}B\leq F_{\rm p,max}$) and (ii) to the maximum variation of the tilt angle between vortices (notice that from Eq.(\ref{eq:ampere}) one has $d\alpha\propto J_{\parallel}$).\cite{bcw} On the other hand, as it was thoroughly discussed in Ref.\onlinecite{gcs}, the model also provides a rule that fixes the {\em trajectory} of the system through the dissipation excursion towards the new equilibrium state. Thus, corresponding to a very sharp transition from the superconducting state to some regime of high losses, one can argue that the final state current density verifies a maximum projection law relative to the transient electric field, i.e.: ${\rm max}\,\,({\bf J}\cdot\hat{\bf E})$ that ensures the fastest return to equilibrium. Notice that,  in 1D situations (infinite slab with a single component applied magnetic field) this is trivially verified because by symmetry one has ${\bf J}\parallel{\bf E}$ and both perpendicular to ${\bf B}$. In other words, one has $J_{\perp}={\rm sgn}(E_{\perp})J_{c\perp}$ with $E_{\perp}$ standing for the component of ${\bf E}$ along the direction ${\bf B}\times ({\bf J}\times{\bf B})$.
In more general configurations, the maximum projection condition is not so simple due to the vectorial character of the problem. Thus, within the DCSM framework, the current density must transit from one state to another that fulfill the conditions $|{\bf J}_{\perp}|\leq J_{c\perp}$ and $|{\bf J}_{\parallel}|\leq J_{c\parallel}$. This may be expressed by a relation of the kind ${\bf J}\in\Delta$ with $\Delta$ having a rectangular section of size $2J_{c\parallel}\times 2J_{c\perp} $ in this case. In summary, the critical state model in general 3D systems is posed by the system of equations
\begin{eqnarray}
\label{eq:maxwell}
\nabla \times {\bf E} &=& - \partial _t{\bf B} \qquad ;\,
\nabla \times {\bf H} = {\bf J}\;\; ({\bf B}=\mu_{0}{\bf H} ) 
\nonumber\\
\nabla\cdot{\bf B} &=& 0 \qquad \qquad;\,{\rm max}\,\,{\bf J}\cdot\hat{\bf E} \,\,\,{\rm with}\,\,\, {\bf J}\in \Delta\, .
\end{eqnarray}
Notice that, as equilibrium magnetization is usually neglected in the critical state regime, ${\bf B}=\mu_{0}{\bf H}$ is used.

The integration of the above system of equations supplemented by appropriate boundary conditions may be cumbersome, so that an alternative formulation has been frequently used, that is fully equivalent, and states the problem in a variational form. It is briefly explained in the forthcoming paragraphs.

\subsection{Variational statement of the critical state problem}

From the mathematical point of view, the above problem is equivalent to the incremental minimization of the functional (field Lagrangian)
\begin{equation}
\label{eq:minprin}
{L}[{\bf H}] \equiv
\int_{\IR ^3}\left[\frac{\mu_0}{2}(\Delta{\bf H})^2 + {\bf p}\cdot(\nabla \times
{\bf H} -{\bf J})\right]d^{3}{\bf r} \, .
\end{equation}

Here, one introduces the variable ${\bf p}$ as a {\em Lagrange multiplier} for enforcing Amp\`ere's law. $\Delta{\bf H}$, on the other hand, represents the magnetic field increment for the time step under consideration. Additionally, the algebraic condition ${\bf J}\in \Delta$ should be fulfilled in the minimization. 

Notice that the Euler--Lagrange equations for this variational problem are 

\begin{eqnarray}
\partial _{\bf p} {\cal L} = &\, 0\, & \Rightarrow \nabla \times {\bf H} - {\bf J}
\nonumber\\ \nonumber\\
\partial _{\bf H}{\cal L} - \partial ^i \left( \frac {\partial {\cal L}}{\partial _i {\bf H}} \right) = &\, 0\, & \Rightarrow \mu _0 
\Delta {\bf H} = - 
\nabla \times {\bf p}
\end{eqnarray}
that identify ${\bf p} \simeq  {\bf E} \Delta t$. From the mathematical point of view, here $\bf J$ is no longer a variable, but plays the role of a parameter to be adjusted in a direct algebraic minimization, i.e.:
\begin{equation}
{\rm max}\,\,(\hat{\bf E}\cdot{\bf J}) \Leftrightarrow {\rm max} \,\,({\bf p}\cdot{\bf J}) 
\end{equation}

As one can see, both Ampere and Faraday's laws are included in the variational formulation, as well as any domain $\Delta$ for the critical current material law. Technically, we emphasize that the shorter is the path step, the better agreement with the standard Maxwell equation formulation. On the other hand, numerical methods for discrete constrained minimization can be used as an alternative to the integration methods for Maxwell's equations, that happends to be very convenient for dealing with 3D problems. 

\subsection{Application: 2D and 3D slab geometry.}

\begin{figure}[t]
\begin{center}
{\includegraphics[width=0.375\textwidth]{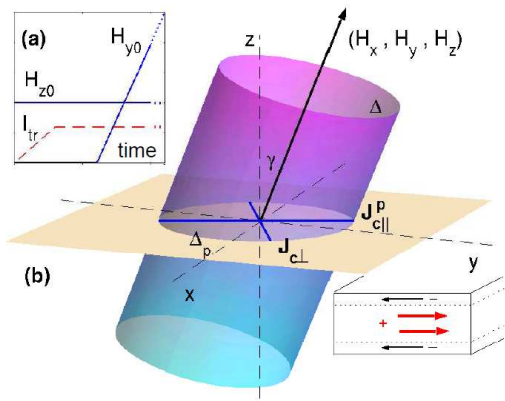}}
\caption{\label{fig_1}(Color online) Better resolution in original paper. Panel (a): magnetic process considered in this
work. A magnetic field $H_{z0}$ is applied perpendicular to the surface of a superconducting slab, that is later
subjected to a transport current along the $y$-axis and to an increasing field $H_{y0}$. Panel (b): the critical current
restriction is represented by a cylindrical region $\Delta$ around the local magnetic field axis (length
$2J_{c\parallel}$ and radius $J_{c\perp}$). $\Delta_p$ is the projection onto the slab ($xy$)-plane. $\gamma$ is the
angle between the field and the $z$-axis. The lower inset depicts an element of the slab and the current backflow.}
\end{center}
\end{figure}

A further advantage of the methodology introduced above is  that the constraint relation ${\bf J}\in\Delta$ allows a pictorial representation that provides a useful tool for the understanding of the current flow structures that arise in the longitudinal configurations. Notice, that in our case (DCSM conditions) $\Delta$ may be depicted by the cylindrical region in Fig.\ref{fig_1}. 

Regarding the specific details about the mathematical technique for obtaining the numerical solution of Eq.(\ref{eq:minprin}), the interested reader is addressed to our Ref.\onlinecite{gcs}. There, we analyzed a number of situations that are easily translated to the study within this work. Here, we will just mention that our proposal consists of transforming the volume integral over the whole space in (\ref{eq:minprin}), into a double integration over the sample's volume. Then, upon discretization, one solves for the distribution of current in a proper set of circuits, under the corresponding constraints for the components of ${\bf J}$ parallel and perpendicular to the local magnetic field. Thus, when one chooses the infinite slab geometry depicted in Fig.\ref{fig_1}, such circuits are naturally defined by a collection of current layers within the $xy$-plane, and each carrying a current density given by $[J_{x}(z_i),J_{y}(z_i)]$, with $z_i$ the position of the layer. Notice that, owing to the planar translational symmetry, a $z$-component of ${\bf J}$ may be ruled out.

We emphasize that the slab geometry enables to study the appearance of the focused physical phenomena with the least mathematical complication, i.e:. this configuration allows to clearly determine the mechanisms related to the negative currents. As a main fact it will be established that, when building the parallel configuration, the response of the superconductor depends on the limitations for the current density established by the depinning threshold $J_{c\perp}$, on the orientation of the local magnetic field, and eventually on the flux cutting restriction $J_{c\parallel}$. This is easily understood at a qualitative level just by glancing at Fig.\ref{fig_1}. The critical current restriction is given by the region $\Delta_{\rm p}$ that is the intersection between the cylinder $\Delta$ and the $xy$-plane, where the current flows. For moderate values of the angle $\gamma$ between the local magnetic field and the $z$-axis, $\Delta_{\rm p}$ is an ellipse of semi-axes $J_{c\perp}$ and $J_{c\parallel}^{\rm p}$ with

\begin{equation}
J_{c\parallel}^{\rm p}={J_{c\perp}}/{\cos{\gamma}}=J_{c\perp}{\sqrt{H_{x}^{2}+H_{y}^{2}+H_{z}^{2}}}/{H_{z}} \, ,
\end{equation}

An increase of the in-plane magnetic field component will result in a tilt of the cylinder, by an increase of the angle $\gamma$. Note in particular that, initially, the maximum value of the in-plane parallel current density, $J_{c\parallel}^{\rm p}$, grows with the angle $\gamma$,  independent of $J_{c\parallel}$ (which is, thus, absent from the theory) until the maximum value $\sqrt{J_{c\perp}^{2}+J_{c\parallel}^{2}}$ is reached.  Then, the ellipse is truncated and eventually would be practically a rectangle of size $2J_{c\parallel}\times 2J_{c\perp}$ when $\gamma \to\pi/2$. Outstandingly, for large values of $\chi\equiv J_{c\parallel}/J_{c\perp}$ (long cylinders), the critical current along the parallel axis, $J_{c\parallel}^{\rm p}$ increases more and more as the weight of $H_{z0}$ decreases and, furthermore, this quantity is always beyond the individual values $J_{c\perp}$ and $J_{c\parallel}$

\section{Simplified analytical model}
\label{secAnalytic}
Here, we show that some of the experimental features that will be obtained later on, from numerical calculations, may be already predicted by a simplified analytical model. Let us consider the excitation process depicted in Fig.\ref{fig_1}  for the particular case $H_{z0}=0 \Rightarrow \gamma=\pi /2$ (the region $\Delta_{\rm p}$ is a rectangle with axes defined by the directions parallel and perpendicular to ${\bf H}$).

\subsection{Governing equations}

Amp\`ere's law takes the following form for the infinite slab geometry considered in this work
\begin{equation}
\label{eq:amperexy}
-\frac{dH_{y}}{dz}=J_{x}
\quad ; \quad
\frac{dH_{x}}{dz}=J_{y} \, .
\end{equation}
On the other hand, following the theory issued in Ref.~\onlinecite{dcsm}, one can show that such expressions may be transformed  into the polar form 
\begin{equation}
\label{eq:ampere}
-H\frac{d\alpha}{dz}=J_{\parallel}^{\rm p}
\quad ; \quad
\frac{dH}{dz}=J_{\perp}^{\rm p}
\end{equation}
with $H=\sqrt{H_{x}^{2}+H_{y}^{2}}$ the modulus of the magnetic field vector, and $\alpha = {\rm atan}(H_y/H_x)$ the angle between such vector and the $x$-axis.

Now, the thresholds of flux depinning and cutting imply the in-plane conditions
\begin{equation}
\label{eq:critical}
|J_{\parallel}^{\rm p}|\leq J_{c\parallel}^{\rm p}(\gamma =\pi /2)=J_{c\parallel}
\quad ; \quad
|J_{\perp}^{\rm p}|\leq J_{c\perp} \, .
\end{equation}

It is apparent that, in general, Eq.(\ref{eq:ampere}) and the conditions in Eq.(\ref{eq:critical}) would not straightforwardly lead to the solution of the problem. Typically, one should also use Faraday's law, either by explicit introduction of the related electric fields (as in Ref. \onlinecite{dcsm}), or by our variational statement. Nevertheless, in this case ($\gamma =\pi/2\Rightarrow J_{c\parallel}^{\rm p}=J_{c\parallel}$), the resolution noticeably simplifies. In fact, for the situation considered, we will have a combination of the cases $J_{\parallel}^{\rm p}=0,\pm J_{c\parallel}$ and $J_{\perp}^{\rm p}=0,J_{c\perp}$ and integration of Eq.(\ref{eq:ampere}) is straightforward. For further mathematical ease, we will also consider $J_{c\parallel}$ and $J_{c\perp}$ to be field independent constants in this work. 

The following normalization, based on the physical parameters that define the problem, will be used: $\vec{\jmath}\equiv {\bf J}/J_{c\perp}$, ${\bf h}\equiv {\bf H}/J_{c\perp}a$ and ${\tt z}\equiv z/a$ ($a$ is the thickness of the slab). The origin of coordinates will be taken at the center of the sample.

Following the notation introduced in Ref.~\onlinecite{dcsm} we will refer to different zones within the sample that are, in brief, macroscopic regions where well defined dissipation mechanisms occur. Inserting our normalized units, there can be T zones, where only flux depinning (transport) occurs (${j}_{\parallel}=0\, ,{j}_{\perp}=\pm 1$), C zones, where only flux cutting occurs (${j}_{\parallel}=\pm\chi\, ,{j}_{\perp}=0$), and CT zones where both transport and cutting occur (${j}_{\parallel}=\pm\chi\, , {j}_{\perp}=\pm 1$). Finally, one will have O zones where neither flux transport nor cutting take place (${j}_{\parallel}=0\, , {j}_{\perp}=0$). Introducing these possibilities in Eqs.(\ref{eq:ampere}) and (\ref{eq:critical}) one gets the following cases for the incremental behavior of the magnetic field in polar components
\begin{eqnarray}
dh=\left\{
\begin{array}{rr}
0\qquad \rm{(O,C)}&
\\
\pm\, dz\;\; \rm{(T,CT)}&
\end{array}
\right.
\!; \;
d\alpha=\left\{
\begin{array}{rr}
0\qquad\qquad \rm{(O,T)}&
\\
\pm\,({\chi}/{h})\,dz\;\; \rm{(C,CT)}&
\end{array}
\right. \, ,
\end{eqnarray}
and all that remains for obtaining the penetration profiles is to solve successively (integrate) for $h$ and $\alpha$ with the corresponding boundary conditions (evolutionary surface values $h_{0},\alpha_{0}$). The case selection has to be made according to Lenz's law. We note in passing that further specification related to the sign is usually included in the notation. Thus, a T$_{+}$ zone will exactly mean $dh = +dz$. 

\subsection{Magnetic process}

\begin{figure}
\begin{center}
{\includegraphics[width=0.45\textwidth]{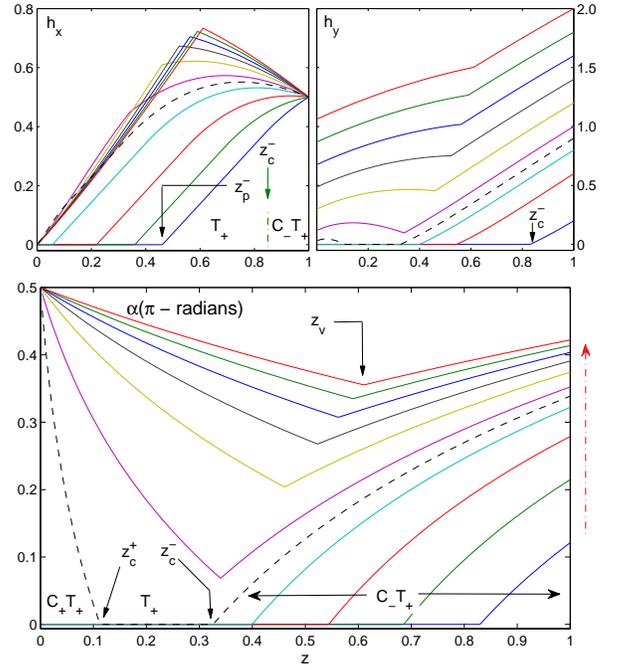}}
\caption{\label{fig_2}(Color online) Penetration of the magnetic field components and rotation angle in the longitudinal transport experiment ($H_{z0}=0$) for a superconducting slab of thickness $2a$, as calculated from Eq.(\ref{eq:ampere}). The zone structure induced by increasing the field $H_{y0}$ is marked upon some of the curves. The dashed line corresponds to the unstable regime (see text). Dimensionless units for $h$ and $z$ are defined in the text.}
\end{center}
\end{figure}

\subsubsection{Application of current}

To start with, the application of the transport current along the $y$-axis produces a T$_{+}$ zone 
\begin{eqnarray}
dh= dz \quad ; \quad d\alpha = 0 
\nonumber\\
{\Downarrow}\qquad\qquad
\\
h=h_{x0}+{\tt z}-1\quad ; \quad \alpha=0 \, ,
\nonumber
\end{eqnarray}
that penetrates from the surface until the point where $h$ equals $0$, i.e.: ${\tt z}_{p0}=1-I_{tr}$. In our units, ${\tt z}_{p0}=0.5$ for $I_{tr}=h_{x0}=0.5$. An O zone appears in the inner region $0<{\tt z}<{\tt z}_{p0}$ as far as $I_{tr}<1$. 

\subsubsection{Application of parallel field: initial steps}

The above situation remains valid until $h_{y0}$ is applied. Then, upon increasing $h_{y0}$, flux line rotation starts on the surface and the perturbation propagates towards the center in the form of a C$_{-}$T$_{+}$ zone defined by 
\begin{eqnarray}
dh= dz \quad ; \quad d\alpha = -\chi/h\, dz\qquad\qquad
\nonumber\\
{\Downarrow}\qquad\qquad\qquad\qquad\qquad
\\
h=h_{x0}+{\tt z}-1\; ; \; \alpha=\alpha_{0}+\chi{\rm ln}[1+({\tt z}-1)/h_{0}] \, ,
\nonumber
\end{eqnarray}
that covers the range ${\tt z}_{c}^{-}<{\tt z}<1$, defined by $\alpha=0\Rightarrow{\tt z}_{c}^{-}=1+h_{0}[{\rm exp}(-\alpha_{0}/\chi)-1]$. The former T$_{+}$ zone is pushed towards the center and occupies the interval ${\tt z}_{p}^{-}<{\tt z}<{\tt z}_{c}^{-}$ with ${\tt z}_{p}^{-}=1-h_{0}$. Finally, an O zone fills the core $0<{\tt z}<{\tt z}_{p}^{-}$.

The upper panes of Fig.\ref{fig_2} sketch the above described structure marked on the cartesian components of the magnetic field vector. The transition points between the different regimes are highlighted.

\subsubsection{Application of parallel field: instability at the center}

The O/T$_{+}$/C$_{-}$T$_{+}$ structure remains valid until the condition ${\tt z}_{p}^{-}=0\Leftrightarrow h_{0}=1$ is reached, i.e.: the modulus of ${\bf h}$ penetrates completely within the interval $0<{\tt z}<1$. Then, the O zone disappears, and a T$_{+}$/C$_{-}$T$_{+}$ structure fills the sample. We emphasize that this configuration becomes unstable owing to the boundary condition $h_{x}({\tt z}=0)=0$ 
that is dictated by the symmetry of $j_y$ around the center. Thus, corresponding to the even behavior of $j_y({\tt z})$, $h_{x}({\tt z})$ must be an odd function.
In physical terms, flux vortices penetrate from the surface with some orientation given by the components of  the vector $(h_{x},h_{y})$. Owing to the critical condition for the penetration of the field $dh/dz = 1$, as soon as the modulus reaches the centre, flux rotation must take place there. This is needed for accommodating the vector to the condition ${\bf h}({\tt z}=0)=(0,h_{y}({\tt z}=0))=(0,h({\tt z}=0))$.
On the other hand, as the angle variation is determined by the value of $J_{c\parallel}$, a jump is induced at the centre, i.e.: $\alpha({\tt z}=0)\to\pi /2$, and the related instability may be visualized by a critical C$_{+}$T$_{+}$/T$_{+}$/C$_{-}$T$_{+}$ profile (dashed line in Fig.~\ref{fig_2}) in which the field angle decreases from its surface value $\alpha_{0}$ to $0$ in the C$_{-}$T$_{+}$ region, then keeps null within the T$_{+}$ zone, and suddenly increases to the value $\pi/2$ in the inner C$_{+}$T$_{+}$ band defined by
\begin{eqnarray}
dh= dz \quad ; \quad d\alpha = \chi/h\, dz\qquad\qquad\quad
\nonumber\\
{\Downarrow}\qquad\qquad\qquad\qquad\qquad
\\
h=h_{x0}+{\tt z}-1\; ; \;  \alpha=\pi/2-\chi{\rm ln}[1+{\tt z}/(h_{0}-1)]\, .
\nonumber
\end{eqnarray}
In fact, a C$_{+}$T$_{+}$/C$_{-}$T$_{+}$ structure is stabilized with the intersection between regions at the point [$\alpha^{{+},{+}}({\tt z}_{\rm v})=\alpha^{{-},{+}}({\tt z}_{\rm v})$] given by ${\tt z}_{\rm v}= 1-h_{0}+\surd{h_{0}(h_{0}-1){\rm exp}[(\pi/2-\alpha_{0})/\chi]}$.
Note that, upon further increasing $h_{y0}$ the point ${\tt z}_{\rm v}$ follows the rule ${\tt z}_{\rm v}(h_{y0}\to\infty)\to (1+h_{x0}/\chi)/2$.  All these features have been marked in the lower pane of Fig.~\ref{fig_2} .

\subsubsection{Physical phenomena}

The previous results allow to identify the following properties as $h_{y0}$ is increased: (i) the appearance of a surface layer with negative transport current density (mind the slope of $h_{x}$ in Fig.(\ref{fig_2}) in view of Eq.(\ref{eq:amperexy})), and (ii) the applied magnetic field {\em re-entry} as related to the inner C$_{+}$T$_{+}$ zone. These features will be confirmed along the forthcoming paragraphs, where the numerical solution of the problem is presented. Additionally, we will show that the inclusion of a third component of the magnetic field ($h_{z0}\neq 0$ in what follows) allows to unveil further details reported in the literature. In particular, the occurrence of the negative current phenomenon at specific locations on the surface of the sample and for a given range of applied magnetic field will be understood within the 3 dimensional scenario.

\section{Numerical results}
\label{secNumeric}

Next, we detail the results obtained numerically for different material laws, as related to the selection of the critical current region $\Delta$. We restrict the plots to the limiting cases $j_{c\parallel}\to\infty$ and $j_{c\parallel} = 1$ (infinite and unit aspect ratio, or also named after T and CT states for obvious reasons). The information of interest for intermediate values is straightforwardly interpolated.
 
From the technical side, we must point out that when minimizing $L$ [see Eq.(\ref{eq:minprin})] a slightly smoothed version of the cylindrical region $\Delta$ has been considered by means of a {\em superelliptic} relation\cite{superellipse} given by 
\begin{equation}
\label{eq:superellipse}
j_{\perp}^{2n}+(j_{\parallel}/\chi)^{2n}\leq 1
\end{equation}
with $n=4$.
This statement performs with a high stability from the numerical point of view.
Figs. \ref{fig_3} and \ref{fig_4} display the main features obtained for the longitudinal transport experiment when the third component of the magnetic field ($h_{z0}$) is incorporated. First, we will analyze the properties of the field (${\bf h}(z)$) and current density (${\vec{\jmath}\,(z)}$) profiles, for a longitudinal configuration built in the fashion described in Fig.(\ref{fig_1}). Fig.(\ref{fig_3}) contains the behavior of these quantities as $h_{y0}$ (applied parallel field) is increased, subsequent to the application of the transport current. This is done for a low and a high value of the perpendicular magnetic field $h_{z0}$.  For the meaning of low and high recall that, along this work, the units are relative to the characteristic penetration field value $H^{*}=J_{c\perp}a$

\subsection{Field and current density penetration profiles}

\begin{figure*}[!]
\begin{center}
{\includegraphics[width=1.0\textwidth]{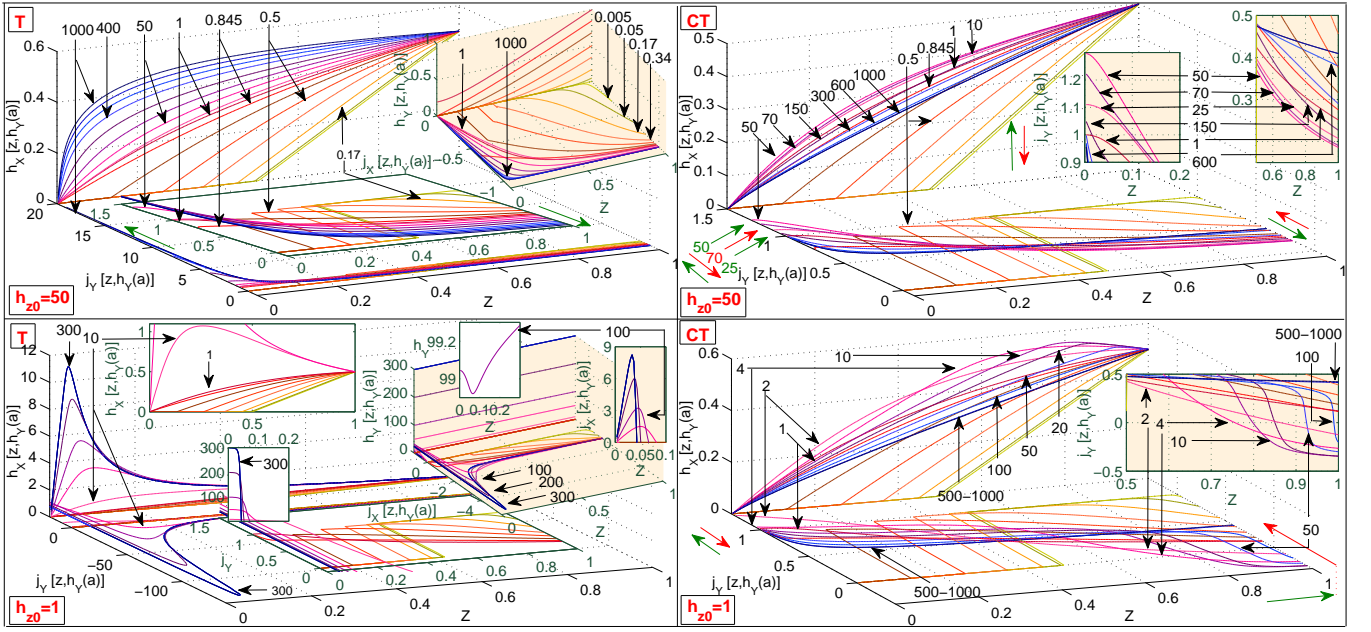}}
\caption{\label{fig_3}(Color online) Profiles of the magnetic field components $h_{x}[z,h_{y}(a)]$ and $h_{y}[z,h_{y}(a)]$, including their corresponding current-density profiles $j_{y}[z,h_{y}(a)]$ and  $j_{x}[z,h_{y}(a)]$ for the \textit{T}-state model (left) and the superelliptic-CT model (right). The magnetic dynamics for high (top) and low (bottom) perpendicular magnetic field $h_{z0}$ is shown. Different scales are used to visualize the intricate behaviour and to emphasize the appearance of negative currents at the surface. The curves are labelled according to the longitudinal magnetic field component at the surface of the slab, $h_{y}(a)\equiv h_{y0}$.}
\end{center}
\end{figure*}

The curves in Fig.(\ref{fig_3}) provide a basic mapping of the physical conditions in which negative currents occur. For completeness and for quantitative purposes, we have included both the field and current density profiles. Nevertheless, they are related by Amp\`ere's law (Eq.(\ref{eq:amperexy})) as one can easily check at qualitative level, i.e.: in the slab geometry $J_{x,y}$ is the slope of $H_{y,x}$ respectively.

Recall that negative values for the transport current density $j_{y}$ are neither obtained for the T or CT states when $h_{z0}$ is high ($h_{z0}\gtrsim 50$) until extreme values of the longitudinal field ($h_{y0}\gtrsim 1000$) are reached. On the contrary, one can early find negative current flow for both cases when $h_{z0}=1$. If $j_{\parallel}$ is unbounded (T states) the ${j}_{y}(z)$ structure becomes rather inhomogeneous as $h_{y0}$ increases and takes the form of a highly positive layer in the center {\em shielded}  by a prominent negative region. When $j_{\parallel}$ is bounded (CT states) one observes a negative layer at the surface that eventually disappears when $h_{y0}$ increases more and more ($h_{y0}>50$).

Some fine structure details are also to be noticed: (i) for the magnetic process under consideration, the {\em partial penetration regime} in which the flux free core progressively shrinks to zero (curves labelled $h_{y0}=0.005\, \cdots\, 0.845$) is practically independent of the critical current model (region) used, (ii) the peaked structure of $j_{y}(z)$ for the T-states at $h_{z0}=1$ (curves labelled $h_{y0}=10 \,\cdots\, 300$) is accompanied by a similar behavior in $j_{x}(z)$ that relates to a subtle magnetic field reentry phenomenon in $h_{y}(z)$ as outlined in the plot, (iii) the negative values of ${j}_{y}(z)$ are obtained for smaller and smaller $h_{y0}$ as $h_{z0}$ also decreases. In fact, negative values can happen even for the partial penetration regime ($h_{y0}\lesssim 0.845$) when $h_{z0}$ tends to $0$, in accordance with the analytical model presented before (Sec.(\ref{secAnalytic})).

\subsection{Experimental quantities}

\begin{figure*}[t]
\begin{center}
{\includegraphics[width=1.0\textwidth]{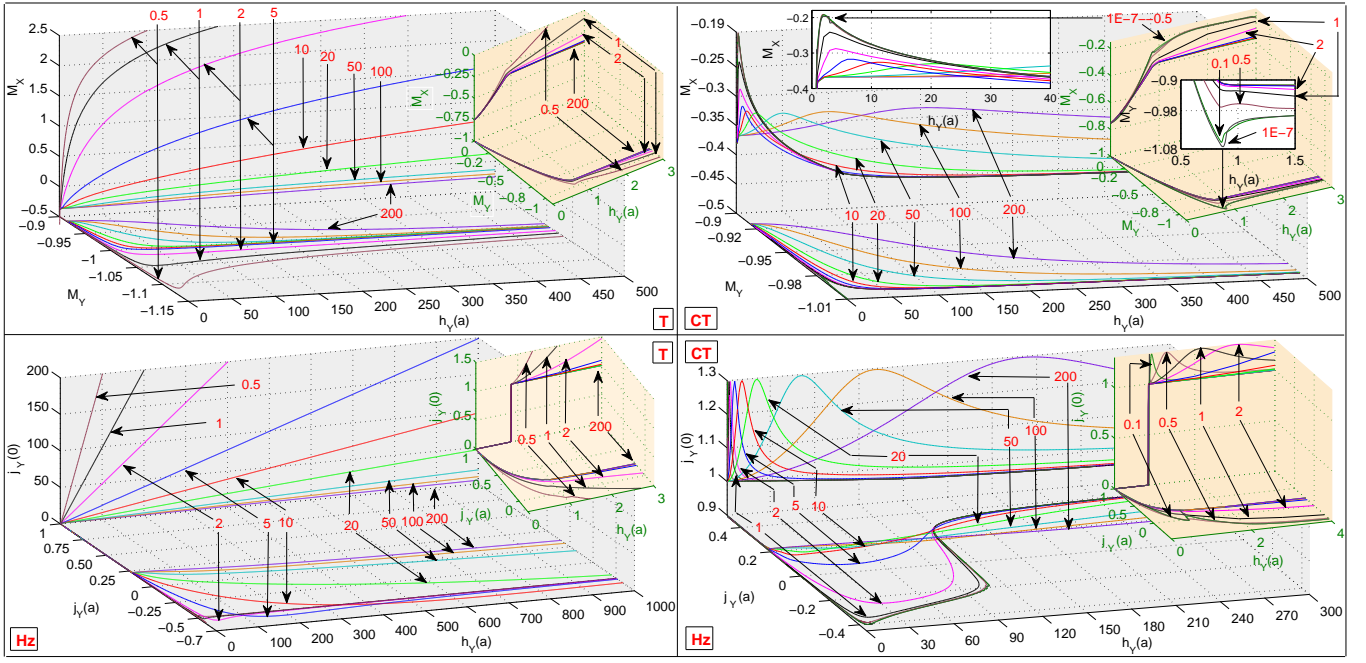}}
\caption{\label{fig_4}(Color online) Top: components of the magnetic moment of the slab as a function of the applied longitudinal magnetic field $h_{y0}=h_{y}(a)$ for the \textit{T}-state model (left) and the superelliptic-CT model (right) with different values of the perpendicular magnetic field component $h_{z0}$ as labelled by arrows on each curve. The magnetization dynamics at low fields $h_{y0}$ is shown in the 3D-insets. Notice the different scales. Bottom: the corresponding evolution of the transport current $j_{y}$ at $(z=0)$ and $(z=a)$.}
\end{center}
\end{figure*}

For a closer connection with real experiments, we have also calculated the sample's magnetic moment ${\bf M}$ as a function of the longitudinal field. The fingerprints of negative current flow will be identified. Fig.~\ref{fig_4} displays the magnetization process of the slab as a function of the applied longitudinal field amplitude $h_{y0}$. ${\bf M}$ (in units of $J_{c\perp}a^{2}$), as well as the transport current density at the center $j_{y}(0)$ and at the surface $j_{y}(a)$ are displayed. 

\begin{figure}[!]
\begin{center}
{\includegraphics[width=.50\textwidth]{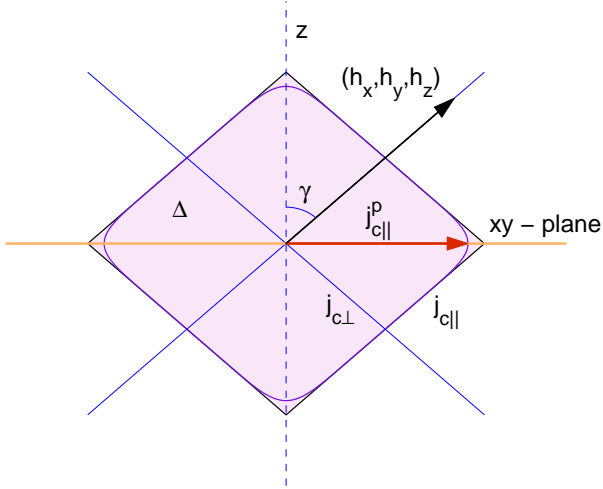}}
\caption{\label{fig_5}(Color online)
Side view of the critical current region $\Delta$ rotated by an angle $\gamma$ that produces a maximal parallel current at the $xy$-plane (this plot is a specific longitudinal section of Fig.\ref{fig_1}). The precise orientation takes place for a definite value of the applied magnetic field $h_{y0}$. The superelliptical region considered in this work has been plotted together with the strictly CT model.}
\end{center}
\end{figure}

Several features are to be identified: 

(i) Unlimited growth of $M_{x}(h_{y0})$ and $j_{y}(0)$ occurs for the T-states, in which $j_{\parallel}$ is unbounded.  On the other hand, the appearance of a peak structure in $M_{x}(h_{y0})$ correlates with a maximum value of the transport current density at the center of the slab for the CT states. The obtained maximum value $j_{y}^{max}(0)= 1.2968$ corresponds to the optimal orientation of the region $\Delta$ in which the biggest distance within the superelliptic hypothesis is reached. Such situation is sketched in Fig.\ref{fig_5} and one may check the numeric result from the expression
\begin{equation}
{\rm max}\; j_{c\parallel}^{p}=(1+\chi^{2n/(n-1)})^{(n-1)/2n} \, ,
\end{equation}
when the choices $\chi=1\, , n=4$ are used. The above formula is obtained from Eq.(\ref{eq:superellipse}) after straightforward calculations. Notice that, as a limiting case, it produces the expected value $2^{1/2}$  for the diagonal of a perfect square (i.e.: $n\to\infty$ in Eq.(\ref{eq:superellipse})). 

(ii) For the unbounded case, in the low $h_{z0}$ regime, the negative current density at the surfaces stabilizes towards the value $j_{y}(a)=-0.5$.

(iii) For the bounded case, and moderate or low $h_{z0}$, the transport current at the surface stabilizes towards the value $j_{y}(a)=0.422$ whether or not it has been negative along the ramp of applied longitudinal field. 

(iv) As a general rule, the smaller the value of $h_{z0}$, the sooner the negative transport current is found. In the CT case, this also increases the range of longitudinal field for which negative values are observed.

\section{Discussion}
\label{secDisc}

Within the previous sections we have displayed a number of cases in which negative transport layers are predicted for type-II superconductors if the longitudinal field configuration (${\bf J}\parallel{\bf B}$) is induced by some external process. Our theoretical investigations allow to identify the following relevant aspects for the appearance of such phenomenon:

(i) the physical mechanisms of {\em flux cutting and depinning}, that may be modelled by the thresholds for the components of ${\bf J}$ parallel and perpendicular to the local magnetic field. In this sense, we have shown that  negative currents are much easier observed for materials in which $J_{c\parallel}$ and $J_{c\perp}$ are independent and $J_{c\parallel}\gg J_{c\perp}$. On the other hand, additional calculations (not displayed) imply that alternative ansatzs as the isotropic model ($J_{\parallel}^2+J_{\perp}^2 \leq J_{c}^2$, i.e.: the difference between the mechanisms responsible for the thresholds $J_{c\parallel}$ and $J_{c\perp}$ are not relevant and the region $\Delta$ is a circle) cannot predict such behavior. However, if some interaction is allowed between the cutting and depinning thresholds [moderately smoothed $J_{\perp}(J_{\parallel})$ relation, i.e.: the region $\Delta$ is a superellipsoid], the negative current flow will occur for some range of conditions.

\begin{figure}[!]
\begin{center}
{\includegraphics[width=.35\textwidth]{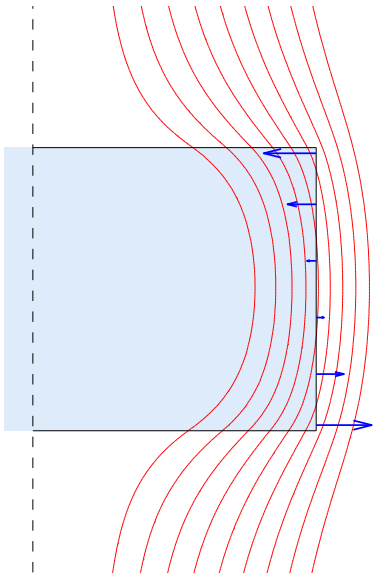}}
\caption{\label{fig_6}(Color online)
Penetration of a magnetic field parallel to the axis of a finite superconducting cylinder (half longitudinal section is shown for symmetry reasons). The component of the magnetic field perpendicular to the lateral surface is visualized by a set of arrows with normalized lengths.The dashed line represents the symmetry axis.}
\end{center}
\end{figure}

(ii) The {\em sequence in which the external excitation components are applied} to the superconductor. Thus, as one could expect from the idea that in the critical state all changes proceed from the surface toward the centre of the sample, negative current structures are enhanced when the applied magnetic field is applied after originally building a subcritical transport profile. Recall that in the situations depicted in Fig.\ref{fig_3}, the initial current flow is compressed more and more, even until compensating negative values are needed at the surface, for maintaining the biased transport current. Along these lines, we should comment that when simulating experiments in which the transport current is applied subsequent to the field, our theory does not predict negative flow values at all. On the contrary, in such cases, what one gets is a {\em compression} of the original field penetration profile, until the increasing transport current leads to dissipation. 

Additional physical considerations can be done so as to cover the full experimental scenario. 
In particular, though our analysis has been done within the infinite slab geometry, one can straightforwardly argue about the extrapolation to real experiments. Thus, the inclusion of the third component of the magnetic field $H_{z0}$ relates to the last relevant aspect:

(iii) The importance of the {\em finite size effects}. Notice that from our numerical calculations, one can predict that negative currents should be more prominent in those regions of the sample where the component of ${\bf H}$ perpendicular to the current layers is less important. This will occur at the central region of the sample's lateral surface, where end effects are minimal. Thus, considering that a real sample in a longitudinal configuration will be typically a rod with field and transport along the axis, the above idea is straightforwardly shown by plotting the penetration of an axial field in a finite cylinder. This has been done in Fig.\ref{fig_6}. The plot shows the distortion of the magnetic field, shielded by the induced supercurrents. Just for visual purposes, we have superimposed the horizontal component of the magnetic field, along the lateral side of the cylinder. It is apparent that the normal component of ${\bf H}$ will be enhanced close to the bases and tend to zero at the central region. Thus, inhomogeneous surface current densities, with negative flow at the mid part should be expected.

\section{Concluding remarks}
\label{secConc}

In this article, we have shown that the counterintuitive effect of negative current flow in type-II superconductors may be predicted and quantified by means of the critical state theory. For restricted situations (infinite slab geometry and fields parallel to the surface), the prediction may even be done within a simplified analytical model. Three dimensional effects may be incorporated by numerical methods when a third component of the magnetic field, perpendicular to the surface of the slab is considered. The analysis of this situation has allowed to conclude that negative transport is enhanced for superconductors in which the flux cutting barrier is much above the depinning value, and at those regions of the sample where the magnetic field is basically parallel to the surface (central part in cylinder geometry).

Extensions of this work are planned along two lines: (i) the actual evaluation of the longitudinal problem in finite length samples (field and transport along the axis of a rod), and (ii) the extrapolation of the current density profiles beyond the critical state threshold, so as to include the dissipation effects.

\section*{Acknowledgment}

This work was supported
by the Spanish CICyT project MAT2008-05983-C03-01 and the DGA grant PI049/08. H. S. Ruiz acknowledges a grant from the Spanish CSIC (JAE program).

%
%


\begin{thebibliography}{99}
\bibitem{superellipse} {Recall that the relation $(x/a)^{2n}+(y/b)^{2n}=1$ represents a family of closed curves (superellipses) with semi-axes $a$ and $b$, ranging from an ellipse to a rectangle as $n$ follows the sequence $n=1,2,3,\dots\infty$ . This was exploited in Ref.\onlinecite{gcs} for comparing different critical state material laws.}
\end{thebibliography}
\end{document}